\documentstyle[epsf,twocolumn,aps,graphics,floats]{revtex}

\begin{document}
\flushbottom
\twocolumn[\hsize\textwidth\columnwidth\hsize\csname@twocolumnfalse\endcsname

\title{\bf Zigzag equilibrium structure in monatomic wires }

\author{
   Daniel S\'anchez-Portal$^1$,
   Emilio Artacho$^2$,
   Javier Junquera$^2$,
   Alberto Garc\'{\i}a$^3$, and
   Jos\'e M. Soler$^{2,4}$
}

\address{
$^1$Department of Physics and Materials Research Laboratory,
    University of Illinois, Urbana, Illinois 61801 \\
$^2$Dep.\ de F\'{\i}sica de la Materia Condensada and
    Inst.\ Nicol\'as Cabrera, C-III,
    Universidad Aut\'onoma, E-28049 Madrid, Spain \\
$^3$Departamento de  F\'{\i}sica Aplicada II,
    Universidad del Pais Vasco, Apdo.\ 644, E-48080 Bilbao, Spain \\
$^4$Department of Physics, Lyman Laboratory, Harvard University,
    Cambridge, MA 02138, USA
}

\date{ECOSS-19 abstract 00675}
\maketitle

\begin{abstract}  
   We have applied first-principles density-functional
calculations to the study of the energetics, and the elastic and
electronic properties of monatomic wires of Au, Cu, K, and Ca
in linear and a planar-zigzag geometries. 
   For Cu and Au wires, the zigzag distortion is favorable even 
when the linear wire is stretched, but this is not observed 
for K and Ca wires.
   In all the cases, the equilibrium structure
is an equilateral zigzag (bond angle of 60$^{\rm o}$).
   Only in the case of Au, the zigzag geometry can also be 
stabilized for an intermediate bond angle of 131$^{\rm o}$.
   The relationship between the bond and wire lengths is 
qualitatively different for the metallic (Au, Cu and, K) 
and semiconducting (Ca) wires.
\end{abstract}

\pacs{Keywords: Density functional calculations, Alkali metals,
      Alkaline earth metals, Copper, Gold, Contacts }

]

   Very recently, using a combination of transmission 
electron microscopy (TEM) and scanning tunneling microscopy, 
Ohnishi {\it et al.}~\cite{Nature1} 
were able to visualize the formation of monatomic gold wires.
   In particular, one of their images shows a wire of 
four gold atoms forming a bridge between two gold tips. 
   The atoms in the wire were spaced at intervals of 3.5-4.0~\AA , 
a distance much larger than that of nearest-neighbors 
in bulk gold (2.88~\AA ). 
   In spite of these large interatomic distances, the wire was 
observed for more than two minutes, showing clearly its stability.
   Observation of gold monatomic chains with a length of four or 
more atoms was independently claimed by other experimental 
group~\cite{Nature2}. 
   In this work, Yanson {\it et al.} do not provide direct structural 
information, but the very long and flexible contacts formed in the 
last plateau of the conductance curves (with a conductance very 
close to one quantum, 2e$^{2}$/h)
were associated to the presence of monatomic chains. 
   These new truly one dimensional structures open a great number of
opportunities to test our understanding of the interactions in solids.
However, the very large interatomic distances found in the experiment 
represent by themselves a puzzle. 
   In ref.~\cite{prl-Au} we already addressed this issue
with the aid of first-principles density-functional
calculations.
   Our estimate of the maximum interatomic distance that the gold 
wires can hold before breaking was $\sim$2.9~\AA, in clear 
disagreement with the commented experimental observations. 
   Several recently published references also report similar
conclusions~\cite{Torres,Okamoto,Hakkinen,DeMaria}.
   The cause for such a discrepancy between theory and experiment
was not clear.
   However, in ref.~\cite{prl-Au} we proposed a simple explanation 
for the puzzle posed by the TEM experiments.
   First, the gold wires were found to exhibit an equilibrium zigzag
structure. 
   Second, the calculated rotational energy barrier was
low enough for the small zigzag wires, supported between gold tips,
to be rotating at room temperature.
   In these conditions, only half of the atoms would be clearly 
visualized, the other half would produce a fuzzy image, 
and the apparent interatomic distance
observed would correspond to that between second-neighbors.

Zigzag structures for monatomic wires have 
already been observed and proposed by several authors. 
Whitman {\em et al.}~\cite{Whitman} observed the formation
of long, isolated zigzag chains when 
Cs was deposited over GaAs(110) and InSb(110) surfaces. 
March and Rubio~\cite{Rubio} also proposed zigzag monatomic chains
as the 
building blocks for highly 
expanded liquid Rb and Cs. A zigzag distortion was also
predicted by Chaplik~\cite{Chaplik} for 
a chain of electrons
held on the surface of liquid helium by the field of a charged filament.

   In this paper we use first-principles density-functional 
calculations to study the stability and electronic properties of 
linear and zigzag wires, for Cu, K, and Ca,
comparing the results with those of Au~\cite{prl-Au}. 
   Our standard density functional theory (DFT)~\cite{DFT}
calculations have been performed with the SIESTA
program~\cite{SIESTA1,SIESTA2}.
   Core electrons are replaced by scalar-relativistic 
norm-conserving pseudopotentials~\cite{JLM}, plus a pseudocore 
to account for non-linear core corrections in the 
exchange-correlation energy~\cite{NLCC}. 
   We used a  doble-$\zeta$ polarized basis 
set of numerical atomic orbitals, obtained from the
solution of the atomic pseudopotentials with an excitation energy 
of 50~meV~\cite{Emilio}.
   Real and reciprocal-space integrations were performed with a
200~Ry-cutoff grid and with 20 inequivalent k-points, which
guarantee a convergence better than 1~meV/atom.
   We present results obtained with the local density approximation 
(LDA)~\cite{LDA} for Au and Cu, and with the generalized gradient 
approximation (GGA)~\cite{PBE} for Ca and K.
   However, we stress that all the results are very similar 
using LDA and GGA.
   We have performed test calculations for the bulk phases of the 
studied materials, obtaining results in good agreement
with experiment and with previous DFT calculations.\footnote{
The calculated equilibrium lattice parameter, 
bulk modulus and, cohesive energy
are: 4.11~\AA, 194~GPa and, 4.40~eV/atom for Au,
3.56~\AA, 210~GPa and, 4.88~eV/atom for Cu,
5.51~\AA, 20.3~GPa and, 2.20~eV/atom for Ca, and 
5.29~\AA, 4.3~GPa and, 0.94~eV/atom for K.}
   In order to minimize the interactions between the wires
and their periodic images, we have allowed 
a distance of 20~\AA\
\begin{figure*}
\epsfxsize=12.0cm\centerline{\epsffile{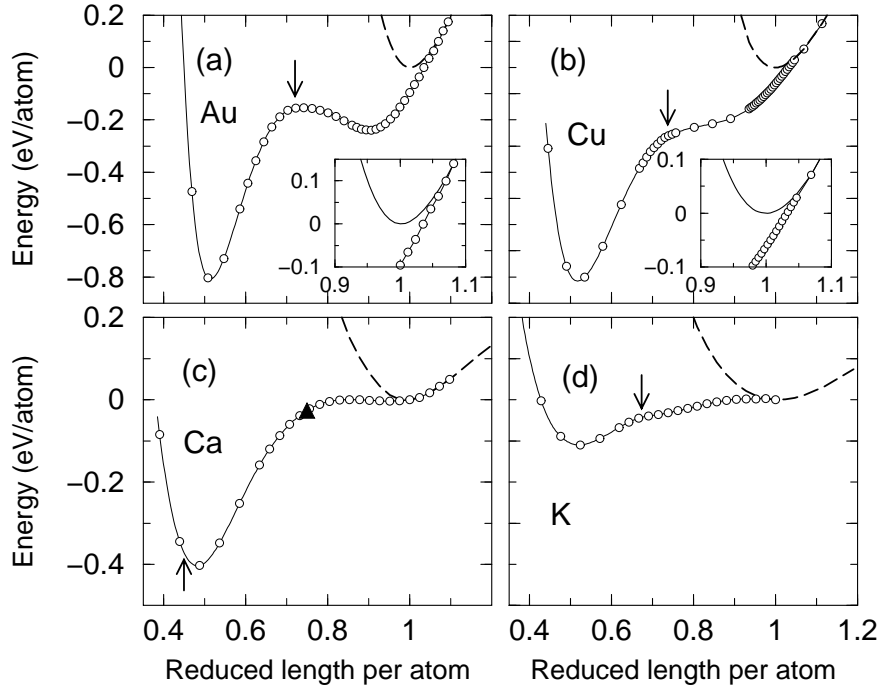}}
\caption[]{
   Energy of monatomic wires of Au (a), Cu (b), Ca (c),
K (d) with a planar zigzag (symbols) and
a linear geometry (dashed lines), as a function of the
length per atom.
   The length $x$ is expressed in units
of the equilibrium bond length of the linear wires,
and the energies are referred to the minimum of the linear wire.
   The arrows indicate points where a conduction band crosses the
Fermi level.
   The triangle shows the length at which the band gap of the
Ca wire moves away from the $\Gamma$ point.
   The insests in (a) and (b) show a zoom
of the energies near the equilibrium of the linear wire.
}
\label{energies}
\end{figure*}
between wires in different cells.

\begin{figure}
\epsfxsize=8.0cm\centerline{\epsffile{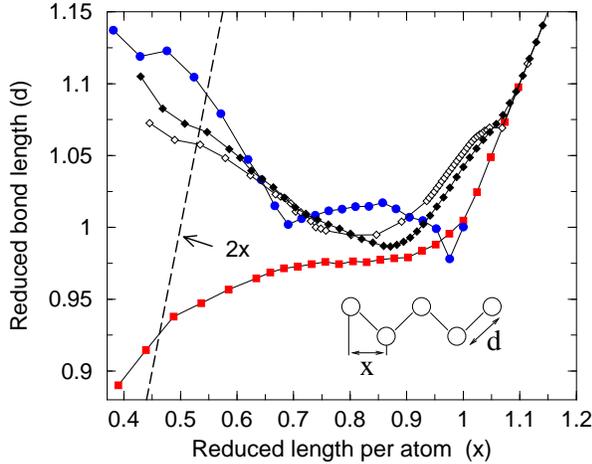}}
\caption[]{
   Evolution of the bond length $d$ for monatomic wires
with a zigzag geometry, as a function of the length per atom $x$.
   All lengths are expressed in units
of the equilibrium bond length of the linear wires.
   Open and solid diamonds stand for Au and Cu, respectively,
circles for K, and squares for Ca.
   The dashed line indicates the distance $2x$
between second-neighbors.
}
\label{distances}
\end{figure}

   Fig~\ref{energies} shows the energies of monatomic
wires with a planar zigzag and linear structure, 
as a function of the wire length, for different elements. 
   To facilitate the comparison, lengths are given in units of 
the linear-chain equilibrium length,
and energies are relative to its minimum. 
   These are: 
2.56~\AA\ and 2.09~eV/atom for Au,
2.25~\AA\ and 2.16~eV/atom for Cu, 
4.10~\AA\ and 0.42~eV/atom for K, and
4.00~\AA\ and 0.98~eV/atom for Ca.
   The evolution of the
bond length in the zigzag structure
can be seen in 
   Fig.~\ref{distances}, where we have again used 
reduced units.
   Hereafter we will use $x$ to refer to the reduced length,
and $d$ for the reduced bond distance
of the wires (see the inset of Fig.~\ref{distances}).

First, we will briefly comment on the data for 
Au, displayed in Fig~\ref{energies}a. 
They present two peculiar behaviors: 
{\it i}) The planar zigzag is more stable 
than the linear geometry, even before the latter
reaches its optimum bond length, and {\it ii}) the zigzag
structure presents two energy minima as a function of the
wire length. While a zigzag structure is naturally expected
under compression, the first point mentioned above 
implies that the zigzag distortion, which increases the
bond length, is favorable even when the linear wire is stretched. 
This is quite surprising in a 
metallic system like gold where, in principle, 
we do not expect any important bond directionality effects. 
Fig.~\ref{energies}b shows 
that this result also holds
for Cu wires. As clearly shown in the insets of 
Fig.~\ref{energies}a and b, in both cases
the linear
configuration is, at its equilibrium 
length ($x$=1),
already $\sim$0.1~eV/atom less stable than the zigzag 
structure. In spite of this energy gain, 
Fig.~\ref{distances} shows that for both materials,
at $x$=1,
the nearest-neighbors distance increases
in $\sim$5~\% when going from the linear
to the zigzag conformation. The behavior of the energy
shown in panels (c) and (d) of Fig.~\ref{energies},
corresponding to Ca and K, is quite different. In those curves
we do
not find any sign (at least within the precision
of the calculation) of the zigzag distortion until the linear
chain starts to be compressed. This is what we could
call a `trivial' zigzag distortion: the energy minimum
of the linear wire approximately becomes an inflection
point for the energy of the zigzag configuration.
This is in fact the kind of behavior 
predicted with, for example, 
a simple pairwise potential. 

All the curves in Fig.~\ref{energies}
present a clear minimum near $x \simeq 0.5$.
This minimum corresponds to a zigzag formed by 
equilateral triangles, where each atom has four nearest-neighbors
at the same distance. In Fig.~\ref{distances} the equilateral
configuration
corresponds to the points where each curve
intersects the dashed line, i.e. where $d=2x$.

\begin{figure}
\epsfxsize=9.0cm\centerline{\epsffile{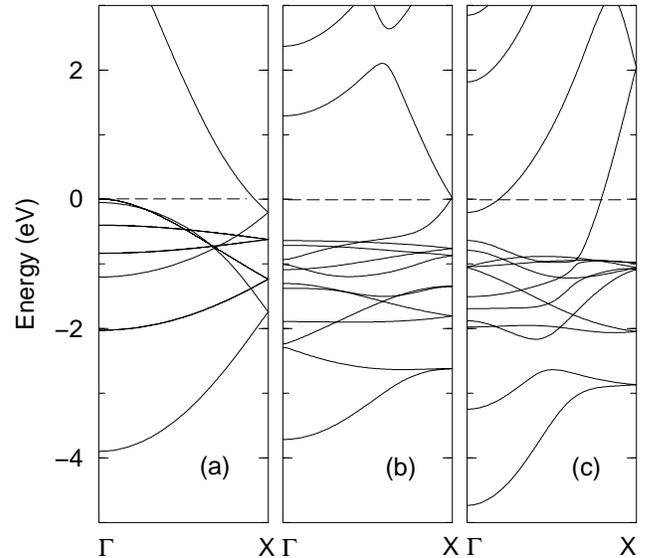}}
\caption[]{
   Band structure of an infinite monatomic copper wire in
a linear geometry of length
2.20~\AA/atom ($x=$0.98) (a),
and a planar zigzag geometry of
1.90~\AA/atom ($x=$0.84) (b), and
1.30~\AA/atom ($x=$0.58) (c).
}
\label{Cu-bands}
\end{figure}

   The second peculiarity of the Au wires
mentioned above, i.e. the presence of two
minima, is not observed for the other species studied. 
   In ref.~\cite{prl-Au} we explained this behavior 
with a simple jellium model, enphasizing the relationship
between the energy vs length behavior, and the crossing of
the Fermi level by different bands.
   For linear wires of Cu and Au with $x<1.1$, a doubly 
degenerated $\pi$ band (with mainly $d$ character) 
rises through the Fermi level (Fig.~\ref{Cu-bands}a).
   In the case of Au linear wires, this was already 
reported in ref.~\cite{prl-Au}, and later confirmed
in ref.~\cite{DeMaria}. 
   The high density of states at the Fermi level, associated 
to this band, destabilizes the linear chain, and is the 
origin of the `anomalous' zigzag behavior of Au and Cu 
(as opposed to the `normal' behavior of  Ca and K), 
which begins when the crossing occurs.

   For $x>0.7$, Au, Cu, and K zigzag-wires have a single
band crossing of the Fermi level (Fig.~\ref{Cu-bands}b),
at the Brillouin zone edge.
   This situation might, in principle, favor a Peirls distortion.
   However, for the wire lenghts studied here, we have not 
found any measurable dimerization.
   As the wire is contracted below $x \simeq 0.7$ 
(the exact positions are indicated by arrows in Fig.~\ref{energies})
a new conduction band crosses the Fermi level. 
   This is shown in Fig.~\ref{Cu-bands}c for the Cu wire.
   For the three elements (Au, Cu and, K), this new band 
crossing precedes a change in the behavior of the energy, and may
be interpreted as the begining of a chemical bond between second 
neighbors.

   The case of the Ca zigzag-wire is different.
   The wire is semiconducting for most
of the length interval studied here, and the bands cross the
Fermi level at a much lower $x \simeq 0.45$ (below the 
equilibrim length) making it metallic.
   At $x \simeq 0.75$ its band structure suffers another
qualitative change: the gap, that was direct at $\Gamma$ for 
larger lengths, begins to move towards the edge of the
two-atom Brillouin zone, still being approximately direct.

   While the energy curves of K and Ca zigzag-wires were relatively 
similar, the behavior of the bond distance, displayed in
Fig.~\ref{distances}, is completely different for metallic
and semiconducting wires. In the case of the metallic wires 
(Au, Cu and, K)
the bond distance increases as the zigzag wire is contracted. 
   This can be understood as a consequence of the
metallic bonding: as the second-neighbour distance decreases,
the first-neighbor distance must increase to maintain
the optimum value of the effective atomic coordination.
   In fact, the calculated behavior is qualitatively 
reproduced by semiempirical metallic potentials like
the glue model~\cite{glue}, 
or the embedded atom method~\cite{embedded},
or even a jellium model where the volume per atom remains fixed.
   However, for the semiconducting wire (Ca), the nearest-neighbor
distance behaves just in the opposite way, decreasing as the wire 
is compressed. 
   It is worth noting here that the predictions of a simple 
pair potential, like the Lennard-Jones~\cite{LJ}, lie exactly 
in between these two regimes, i.e. the nearest-neighbors
distance would remain almost constant 
once the wire assumes the zigzag configuration.
   In Fig.~\ref{distances}, it is also clearly visible 
the abrupt change in the behavior of the bond length of K wires
induced by the band crossing at $x \simeq 0.7$.

   In conclusion, we have applied first-principles density-functional
calculations to study the energetics, and the elastic and 
electronic properties of monatomic wires of different elements,
in a linear and a planar zigzag geometry. 
   Surprisingly, for Cu and Au wires, a zigzag distortion is 
favorable even when the linear wire is stretched. 
   This phenomenon is not observed
for K and Ca wires, and might be related with the presence 
of $d$ bands (with a high density of states associated) 
at the Fermi level for the linear conformation of Au and Cu wires.
   For all the studied materials, the equilibrium structure
is an equilateral zigzag (bond angle of 60$^{\rm o}$). 
   Only in the case of Au, the zigzag geometry can also be 
stabilized for an intermidiate bond angle of 131$^{\rm o}$.
   The behavior of the nearest-neighbor bond length 
is qualitatively different in the metallic (Au, Cu and, K) and 
semiconducting wires (Ca). 

   D.S.P. is grateful to Richard M. Martin for advice and support. 
This work was supported by Grants No. DOE~8371494, 
DEFG~02/96/ER~45439.
   J.M.S. acknowleges a sabbatical fellowship from Spain's SEUID.

\end{document}